\documentclass[aps,eprint,twocolumn,groupedaddress,superscriptaddress]{revtex4}

\usepackage{amsmath}
\usepackage{amssymb}
\usepackage{graphicx}

\newcommand{\nhat}{\hat{n}}
\newcommand{\hhat}{\hat{h}}

\begin{document}


\title{Anomalous Expansion of Attractively Interacting Fermionic Atoms\\
 in an Optical Lattice}

\author{Lucia Hackerm\"uller}
\author{Ulrich Schneider}
\author{Maria Moreno-Cardoner}
\affiliation{Institut f\"ur Physik, Johannes Gutenberg-Universit\"at, 55099 Mainz, Germany}
\author{Takuya Kitagawa}
\affiliation{Physics Department, Harvard University, Cambridge MA 02138, USA}
\author{Thorsten Best}
\author{Sebastian Will}
\affiliation{Institut f\"ur Physik, Johannes Gutenberg-Universit\"at, 55099 Mainz, Germany}
\author{Eugene Demler}
\affiliation{Physics Department, Harvard University, Cambridge MA 02138, USA}
\author{Ehud Altman}
\affiliation{Condensed Matter Physics Department, Weizmann Institute of Science, Rehovot 76100, Israel}
\author{Immanuel Bloch}
\author{Bel\'{e}n Paredes}
\affiliation{Institut f\"ur Physik, Johannes Gutenberg-Universit\"at, 55099 Mainz, Germany}

\date{\today}


\begin{abstract}
Strong correlations can dramatically modify the thermodynamics of a quantum many-particle system. Especially intriguing behaviour can appear when the system adiabatically enters a strongly correlated regime, for the interplay between entropy and strong interactions can lead to  counterintuitive effects. A well known example is the so-called Pomeranchuk effect, occurring when liquid $^3$He is adiabatically compressed towards its crystalline phase. Here, we report on a novel anomalous, isentropic effect in a spin mixture of attractively interacting fermionic atoms in an optical lattice. As we adiabatically increase the attraction between the atoms we observe that the gas, instead of contracting, anomalously expands. This expansion results from the combination of two effects induced by pair formation in a lattice potential: the suppression of quantum fluctuations as the attraction increases, which leads to a dominant role of entropy,  and the progressive loss of the spin degree of freedom, which forces the gas to excite additional orbital degrees of freedom and expand to outer regions of the trap in order to maintain the entropy. The unexpected thermodynamics we observe reveal fundamentally distinctive features of pairing in the fermionic Hubbard model.
\end{abstract}

\maketitle


\subsection{Introduction}
The striking consequences of strong correlations in many-body quantum systems are at the frontier of current research. Typically, interest is devoted to the unusual properties of ground states or low-lying excitations, which range from exotic types of order to unconventional quasiparticle statistics \cite{Auerbachbook,Wen:2004}. But strong correlations can also severely alter the thermodynamics of a quantum system, leading to fascinating finite temperature effects. Especially surprising behaviour can arise when the system adiabatically enters a strongly correlated phase, as the emerging correlations can imply a substantial redistribution of entropy.


A well known example of this type of phenomena is the so-called Pomeranchuk effect \cite{Pomeranchuk:1950,Richardson:1997}, which occurs in the liquid to solid transition of $^3$He. Since the solid, through its randomly oriented spins, is more disordered than the liquid, it turns out that when the liquid is adiabatically squeezed, it freezes into a solid by, astoundingly, absorbing heat. 


Other examples of anomalous behaviour due to the combination of finite entropy and strong correlations have been rarely observed in nature, usually because interactions in typical strongly correlated systems can hardly be tuned. Recently, the extraordinary progress in the control and manipulation of neutral atoms in optical lattices \cite{Jaksch:2005,Lewenstein:2007,Bloch:2008c} has added a valuable degree of freedom to the investigation of strongly correlated systems. By varying a collection of parameters, like the scattering length, the lattice depth or the external confinement, it is possible to adiabatically bring a non-interacting gas of bosonic \cite{Fisher:1989,Jaksch:1998,Greiner:2002a,Paredes:2004,Kinoshita:2004,Stoferle:2004,Spielman:2007,Mun:2007,Fallani:2007,Spielman:2008,Guarrera:2008} or fermionic atoms \cite{Chin:2006,Strohmaier:2007,Jordens:2008,Schneider:2008a} into a regime of strong correlations. Here, we report on the experimental observation and the theoretical prediction of a novel instance of an anomalous isentropic effect for a spin mixture of attractively interacting fermionic atoms in an optical lattice.


Our observation moreover constitutes a step towards the study of pairing and superfluidity of fermionic atoms in optical lattices. In the continuum, fermionic atoms have been the subject of breakthrough dipole trap experiments studying the BCS-BEC crossover \cite{Regal:2004a,Zwierlein:2004,Bartenstein:2004b,Bourdel:2004}. In optical lattices, one of the major challenges is the realization of low-temperature quantum phases within a single band fermionic Hubbard model, which could provide critical insight into the origin of high temperature superconductivity in cuprates \cite{Lee:2006,Hofstetter:2002}. In this context, the atomic quantum simulation of the attractive Hubbard model has found special interest \cite{Toschi:2005,Chien:2008,Paiva:2009,Ho:2009}. On the one hand, it allows the investigation of pairing in a single band Hubbard model and gives access to the intriguing preformed pair or pseudogap regime \cite{Toschi:2005,Chien:2008,Paiva:2009}. On the other hand, it could serve as an alternative route, more accessible to current experiments, to study the physics of the repulsive Hubbard model \cite{Ho:2009}. The work we present here contains the first experimental study of the low entropy thermodynamics of an increasingly attractive fermionic spin mixture in the lowest band of an optical lattice. Our theoretical explanation of the unusual behaviour we observe reveals  the dramatically different consequences of pair formation in the single band fermionic Hubbard model with respect to the continuum case.

\begin{figure}
\begin{center}
\includegraphics[width=0.9\columnwidth]{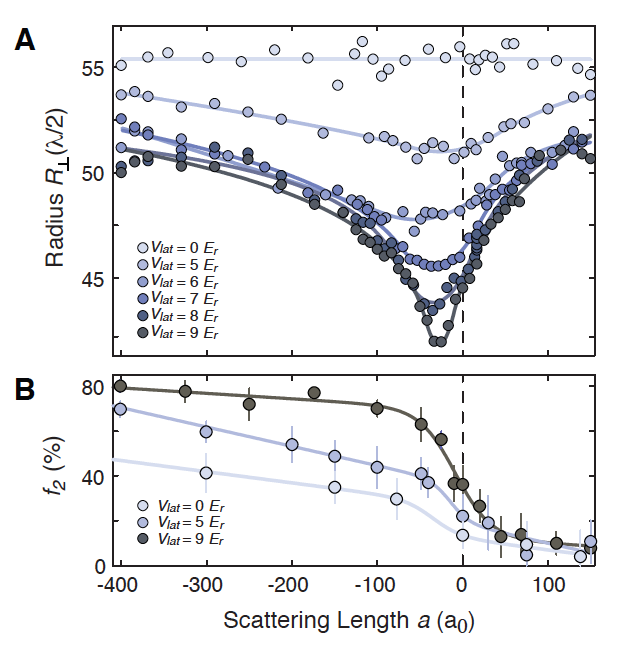}
\end{center}
\caption{{\bf Experimental observation of anomalous expansion.} 
{\bf A}, measured cloud size $R_\bot$ and {\bf B}, fraction of particles on doubly occupied sites $f_2$ (doublon fraction) versus scattering length for different lattice depths ($0$ to $9\,E_r$). 
Dots in {\bf A} correspond to a running average over three experimental shots. Dots in {\bf B} are averages over at least five consecutive measurements, with the standard deviation plotted as the error bar. Lines are guides to the eye. The data have been taken in a fixed external dipole trap with $\omega_\perp=2\pi \times 25$\,Hz and aspect ratio $\gamma \approx 4$, at a fixed temperature prior to loading of the lattice of $T/T_F=0.15(3)$ (see appendix).
The data show an initial contraction of the gas for weak attractive interactions followed by an expansion for strong interactions. As the gas expands, the doublon fraction continues to increase.\label{fig:Experimental}}
\end{figure}

For a spin mixture of trapped fermions, one would expect that increasing the attractive interactions between spin components will compress the gas. 
In order to decrease its energy the system increases its density, the attraction playing the role of an effective increase in the confining potential. This behaviour has been observed for fermionic atoms in recent dipole trap based experiments \cite{Bartenstein:2004b}. 
Here, we show that completely different physics appears when the gas is loaded into a lattice potential: while the system contracts for weak attractive interactions, it reaches a minimum size for a certain finite interaction strength and then starts to increase in size, becoming even larger than the  non-interacting gas (Fig.~\ref{fig:Experimental}). If, in analogy to the usual volume compressibility\cite{Schneider:2008a}, we define the {\em interaction compressibility} as the change in size due to a change in interaction strength, the system exhibits an anomalous, negative interaction compressibility for strong attractive interactions.

This counterintuitive crossover from contraction to expansion is the consequence of the progressive suppression of quantum fluctuations that pairing induces in a lattice potential. As the attractive interaction increases, the system turns into a gas of hard-core pairs, in which the two fermions with opposite spins are tightly bound at the same lattice site.  These on-site pairs can only move through the lattice via virtually breaking up \cite{Auerbachbook,Foelling:2007} and, in contrast to the continuum case, their kinetic energy vanishes as their binding energy increases.

The quenching of quantum fluctuations as the attraction increases has two dramatic effects. On the one hand, it effectively enhances the role of entropy, which becomes increasingly dominant over energy as the attraction gets stronger. On the other hand, it makes the bosonic pairs acquire a fermionic character. Since Pauli principle does not allow two on-site pairs to meet at the same lattice site, the more and more spatially localized pairs eventually behave as spinless fermions. 

Unlike what happens in the continuum, where increasing the attraction between the spin components converts the non-interacting fermionic gas into a gas of pairs that Bose condense, pairing in a lattice potential brings the system back to a fermionic gas, which has however lost its spin degree of freedom. The progressive loss of the spin degree of freedom gives rise to a redistribution of entropy from spin to orbital degrees of freedom.  In order to maintain the entropy, the system is excited to outer regions of the trap and forced to expand (see Fig.~\ref{fig:Schematic}B-C).


\subsection{The System}
\begin{figure}
\begin{center}
	\includegraphics[width=0.9\columnwidth]{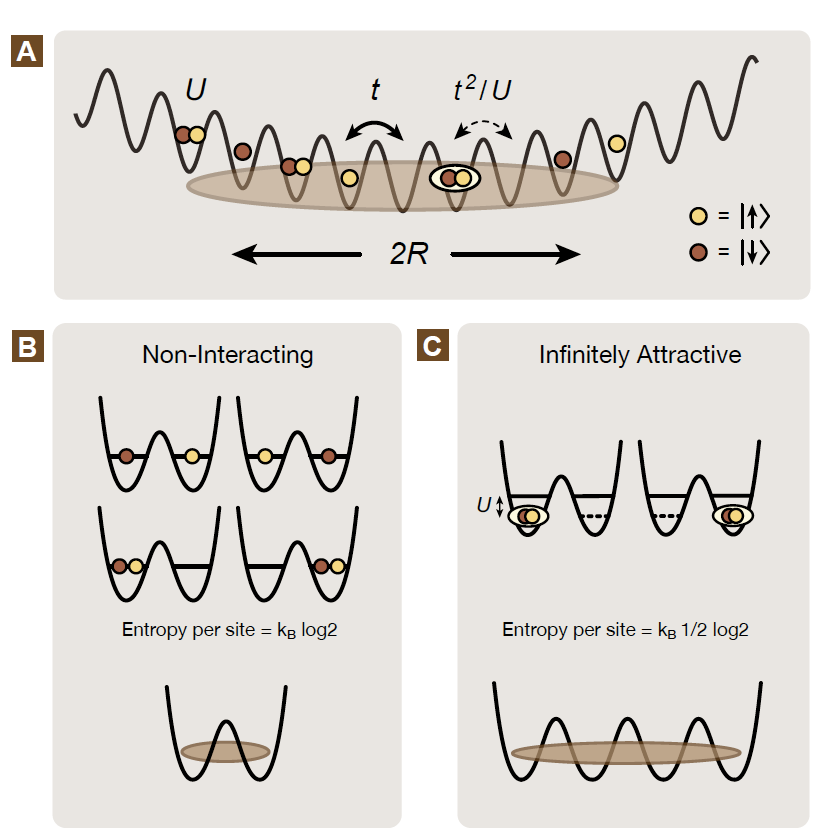}
\end{center}
\caption{{\bf Attractively interacting fermionic spin mixture in an optical lattice}.	
{\bf A} The system size $R$ is determined by the interplay among entropy and the different energy scales: the interaction $U$, the tunneling amplitude $t$ and the confinement energy. For strong attractive interaction the formation of on-site pairs forces the system to expand. {\bf B, C}, Schematic illustration of the anomalous expansion effect with a zero-tunneling two-particle model. The entropy that can be stored per site is reduced as the system evolves from {\bf B}, the non-interacting regime, with $4$ possible configurations, to {\bf C}, the infinitely attractive regime, where particles are tightly paired and only $2$ configurations are available.\label{fig:Schematic}}
\end{figure}

We consider an attractively interacting spin mixture of fermionic atoms loaded into the lowest band of a three dimensional optical lattice. Its physics can be described by a Hubbard Hamiltonian with an additional harmonic confining potential:
\begin{equation}
\hat{H}=
-t
\sum_{\langle \ell , \ell'\rangle \sigma}
c^{\dagger}_{\ell\sigma}
c^{\,}_{\ell'\sigma}+
U\sum_{\ell}
\nhat_{\ell \uparrow}\nhat_{\ell \downarrow}
+E_c
\sum_{\ell \sigma}
r_\ell^2
\nhat_{\ell \sigma}.
\label{Hubbard}
\end{equation}
Here $c_{\ell \sigma}$($c^{\dagger}_{\ell \sigma}$) and $\hat{n}_{\ell \sigma}$ are, respectively, the fermionic destruction (creation)
operator and the particle number operator at lattice site $\ell=(x,y,z)$ and spin state $\sigma \in \{\uparrow, \downarrow \}$. We consider the case of an unpolarized system, with $N_{\uparrow}=N_{\downarrow}=N/2$. The Hamiltonian (\ref{Hubbard}) consists of three competing terms. The first term accounts for the kinetic energy of the system, which is characterized by the hopping amplitude $t$ between neighboring lattice sites, while the second term describes the attractive on-site interaction $U<0$ between atoms 
with opposite spin (see Fig.~\ref{fig:Schematic}A).
The last term takes into account the confinement energy due to the external anisotropic harmonic potential. The characteristic energy $E_c= V_c r_c^2$ is the mean potential energy per particle and spin state of a maximally packed state at the bottom of the trap, 
where $r_c^2d^2$ is the corresponding mean squared radius, $V_c=\frac{1}{2}m\omega_\perp^2d^2$, 
$\omega_\perp=\omega_x=\omega_y= \omega_z/\gamma$ is the horizontal trap frequency, and $d$ the lattice constant. 
The squared radius at site $\ell$ is $r_\ell^2=1/r_c^2\left(x^2+y^2+\gamma^2z^2\right)$.


We want to study the size behaviour of the system when adiabatically entering the regime of dominating attractive interaction $U$. In order to characterize the size we define the radius $R$ in the form:
\begin{equation}
R^2=\frac{1}{N_\sigma}\sum_\ell  r_\ell ^2 n_{\ell},
\end{equation}
where $n_{\ell}=\langle \nhat_{\ell \sigma} \rangle$.
With this definition the average potential energy per particle and spin state is $E_cR^2$ and the radius of the maximally packed state is equal to one.
As a measure of the change in size in response to an adiabatic change of the interaction strength we define the interaction compressibility  
\begin{equation}
\kappa_i=\frac{\partial R^2}{\partial U} \biggr \rvert_{S}.
\end{equation}
This is analogous to the volume compressibility $\kappa_c=-\partial R^2/\partial E_c \rvert_{S}$, which describes the behaviour of the radius with external confinement \cite{Schneider:2008a}.

At zero temperature the size of the system monotonically decreases with increasing attractive interaction and $\kappa_i$ is always positive (see Fig.~\ref{fig:RadiusCompetition}A). As for the continuum case, this behaviour can be qualitatively understood within a mean-field picture, in which the increasing attractive interaction plays the role of an additional trapping potential that effectively compresses the gas.
What is special of the lattice system is the progressive quenching of kinetic energy as the attractive interaction increases, which gives rise to a saturation of the system size in the strongly interacting regime. As the attraction gets stronger, fermions with opposite spins get bound forming on-site pairs, which can only tunnel via second-order processes of suppressed amplitude $t^2/U$ (see ref.˜\cite{Auerbachbook,Foelling:2007}). In the limit of a dominating attractive interaction, pairs get localized at lattice sites, and are obliged by Pauli principle to singly occupy the lattice sites at the bottom of the trap. Once this maximally packed state with $R=1$ is reached, Pauli blocking avoids further compression of the system with increasing attractive interaction and  $\kappa_i$ tends to zero.

We show below how this suppression of quantum fluctuations leads to unexpected size behaviour at finite entropy. Since the desire of minimizing the energy influences less and less the system size, entropy acquires a dominant role as the attraction increases.


\subsection{Experimental Observation}
In the experiment an equal mixture of quantum degenerate fermionic $^{40}$K atoms in the two hyperfine states 
$\vert F, m_F \rangle=\vert \frac {9}{2}, -\frac {9}{2}\rangle \vert \equiv \vert\!\! \downarrow \rangle$ and 
$\vert \frac {9}{2}, -\frac {7}{2}\rangle \equiv \vert \!\!\uparrow \rangle$ is used. 
By overlapping two orthogonally propagating lasers with elliptical beam shapes, a pancake-shaped dipole trap with an aspect ratio $\gamma \approx 4$ is formed. Using evaporative cooling in this trap, temperatures down to $T/T_F = 0.12(3)$ with $1.4-1.8\times10^5$ atoms per spin state are reached. The combination of a red detuned dipole trap ($\lambda_{dip}=1030$\,nm) and a blue detuned optical lattice ($\lambda_{lat}=738$\,nm) with simple cubic geometry allows an independent control of the confinement energy $E_c$ and the tunneling $t$. By means of a Feshbach resonance located at $202.1$G (see ref.˜\cite{Regal:2003b}), the scattering length between the two spin states can be varied, and thereby the onsite interaction energy $U$ can be tuned at constant tunneling. Negative scattering lengths up to $a=-400\,a_0$ can be reached, where a further approach to the Feshbach resonance starts to be hindered by enhanced losses, heating and non-adiabatic effects in the lattice (see appendix).

After evaporation, the dipole trap depth is ramped in $100$\,ms to the desired value of the external confinement ($\omega_\perp=2\pi \times 20-70$\,Hz) and the magnetic field is adjusted to set the scattering length. Subsequently, the optical lattice is increased to a potential depth between $V_{lat}=0-9E_r$  with a ramp rate of $7$\,ms$/E_r$ (see appendix), $E_r=h^2/(2m\lambda_{lat}^2)$ being the recoil energy.

In order to measure the size of the system, an in-situ image of the cloud along the short axis of the trap is taken using phase-contrast imaging \cite{Andrews:1996a}. From this image the integrated perpendicular radius $R_\perp=\sqrt{\langle x^2+y^2 \rangle}$ is obtained via an adapted Fermi-Dirac fit (see appendix).
The behaviour of the system size with increasing scattering length is analyzed for various lattice depths (Fig.~\ref{fig:Experimental}A) and various confinements (Fig.~\ref{fig:Comparison}A). The data obtained shows a contraction of the gas for weak attractive interactions followed by an anomalous expansion for interactions larger than a critical value, which typically corresponds to a scattering length $\vert a \vert \approx 20-40\,a_0$. 
Additionally, the fraction of atoms sitting on doubly occupied sites (doublon fraction) is measured via conversion into molecules \cite{Regal:2003b, Stoferle:2006, Schneider:2008a} (see appendix), showing a steep increase as the interaction becomes attractive. The number of doublons surprisingly continues increasing while the gas expands, saturating close to $80\%$ for strong interactions and deep lattices. This high doublon fraction, at a density substantially lower than 2 atoms per site, indicates that the system is in a preformed pair regime \cite{Paiva:2009}.
In the absence of the lattice, we find that the anomalous expansion disappears and the size of the cloud remains constant, while the doublon fraction can still increase to above $>40\%$ when the free atom cloud is projected into the lattice (see appendix). The absence of expansion in the continuum system has also been observed in recent dipole trap based experiments carried out at comparable (within a factor of two) entropies \cite{Bartenstein:2004b}.

\subsection{Theoretical Prediction}
We start by showing that in the strongly interacting regime the system exhibits a negative interaction compressibility. This is done by considering the zero tunneling limit, in which quantum fluctuations are completely suppressed and the effect of entropy can be isolated. Based on this knowledge, we then show how, at finite tunneling, a competition between energy minimization and entropy conservation leads to a transition from positive to negative compressibility as the attraction increases.

\begin{figure}
\begin{center}
		\includegraphics[width=0.9\columnwidth]{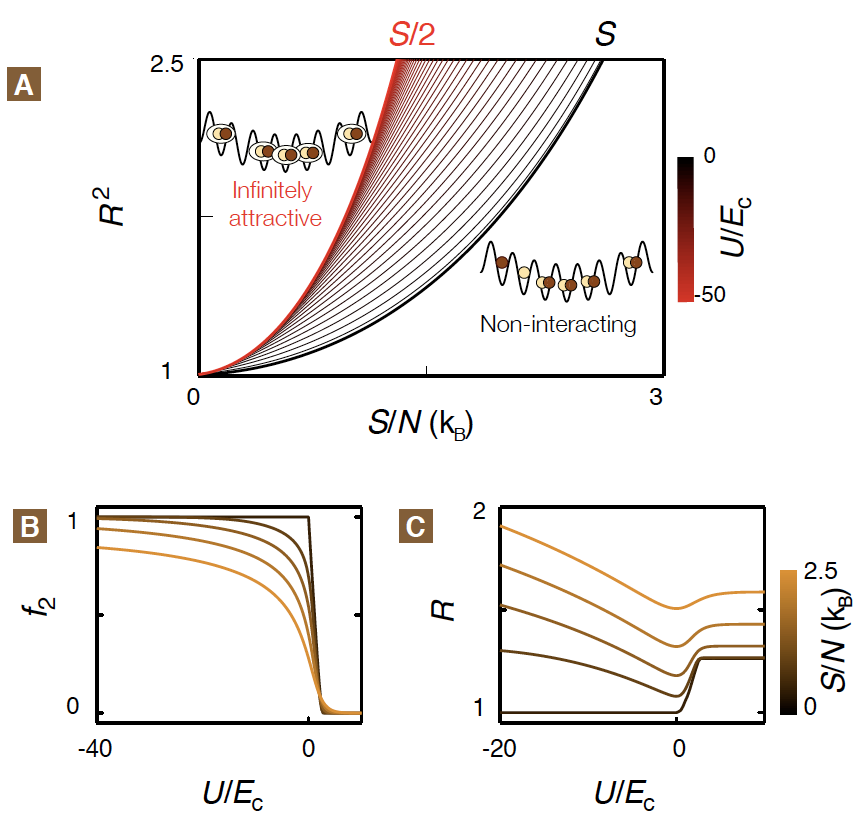}
\end{center}
\caption{{\bf Origin of negative compressibility at zero tunneling.} Numerical exact calculation at zero tunneling for a three dimensional system with $N_\sigma=7.5\times 10^3$. {\bf A}, reduction of entropy density for increasing attraction. The squared radius is plotted versus entropy for different attractive interaction strengths.  For a fixed size the corresponding entropy monotonically decreases as the attractive interaction increases from $U=0$ (black curve) to $U=-\infty$ (red curve). For a fixed non-vanishing entropy the radius increases as the attraction increases. {\bf B}, fraction of particles on doubly occupied sites (doublons) and {\bf C}, radius versus interaction strength at different fixed entropies.  
\label{fig:ZeroTunneling}}
\end{figure}

{\em Zero tunneling limit. Negative compressibility}. 
At zero tunneling and zero temperature the system is in a maximally packed state for any attractive interaction strength and the interaction compressibility vanishes. 
For finite entropy the size behaviour is then only determined by the change in entropy density as the interaction increases, which can be calculated exactly for any interaction strength (see appendix).
Since the Hamiltonian is a sum of local Hamiltonians at each site,
\begin{equation}
\hhat_\ell=U\nhat_{\ell \uparrow}\nhat_{\ell\downarrow}+E_c r_\ell^2(\nhat_{\ell \uparrow}+\nhat_{\ell\downarrow}),
\end{equation}
the problem factorizes into local on-site problems characterized by the probabilities for zero, single and double site occupation. 
As interaction increases from zero to infinitely attractive, single occupation is progressively suppressed and the system evolves from a gas of non-interacting fermions with spin, and local entropy 
$s_\ell=-2 [n_\ell \log n_\ell + (1-n_\ell) \log (1-n_\ell)]$, 
to a system of on-site pairs. 
In contrast to pairs in the continuum which can Bose condense in the same quantum state, these hard-core bosons occupy lattice sites according to Fermi statistics and have local entropy $s_\ell=-[n_\ell \log n_\ell + (1-n_\ell) \log (1-n_\ell)]$, as if they were fermions without spin. For the same density (the same radius) the entropy is thus exactly reduced by a factor of two.
This reduction in the entropy density is illustrated in Fig.~\ref{fig:ZeroTunneling}A, which shows the relation between the entropy and the size of the system for increasing attractive interactions. 
For a fixed entropy this reduction forces the system to expand as the attraction increases, exhibiting a negative interaction compressibility for any non-vanishing attraction and entropy  (Fig.~\ref{fig:ZeroTunneling}C).
In contrast to the mean-field picture, volume compressibility and interaction compressibility have here opposite signs.

\begin{figure}
\begin{center}
\includegraphics[width=0.9\columnwidth]{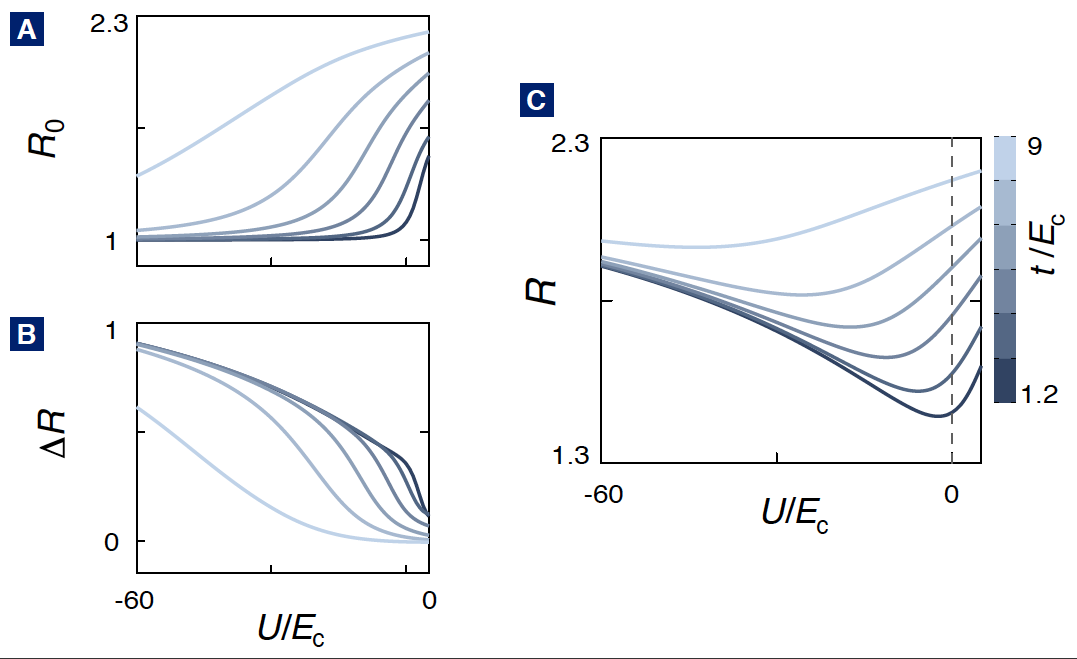}
\end{center}

\caption{{\bf Competition between energy minimization and entropy accommodation at finite tunneling}. Radius versus attractive interaction at constant entropy: exact diagonalization of a one-dimensional system with $6$ particles and $7$ sites. {\bf A}, the radius at zero temperature, $R_0$, decreases with attraction to minimize the total energy. {\bf B}, the change of the radius due to entropy $\Delta R$, however, increases with attraction as a consequence of the gradual loss of the spin degree of freedom. {\bf C}, the resulting radius $R=R_0+\Delta R$ reaches a minimum value for a certain interaction strength where the entropy effect starts dominating the energy effect. The different curves correspond to different ratios of $t/E_c$. Entropy per particle is $S/(k_BN)=0.59$.\label{fig:RadiusCompetition}}
\end{figure}


{\em Finite tunneling. Transition from positive to negative compressibility}.
For finite tunneling and at zero temperature the radius decreases with increasing attractive interaction and the system exhibits a positive interaction compressibility. At finite entropy this radius, $R_0$, has to be increased by an amount $\Delta R$ in order to accommodate the given entropy:
\begin{equation}
R=R_0+\Delta R(S).
\end{equation}
In contrast to $R_0$, $\Delta R$ increases as the attraction becomes larger, for the progressive loss of the spin degree of freedom implies a reduction of the average entropy that can be stored per lattice site.
Energy minimization and entropy conservation thus compete for the sign of the compressibility.

For weak attractive interaction, we expect the behaviour of the radius to be dominated by the zero entropy radius, and the system to exhibit a positive interaction compressibility.
For strong enough attractive interaction, however, where the influence of energy becomes negligible, the reduction of entropy density should dominate and the system should  increase its size, exhibiting a negative interaction compressibility.

\begin{figure}
\begin{center}
	\includegraphics[width=0.95\columnwidth]{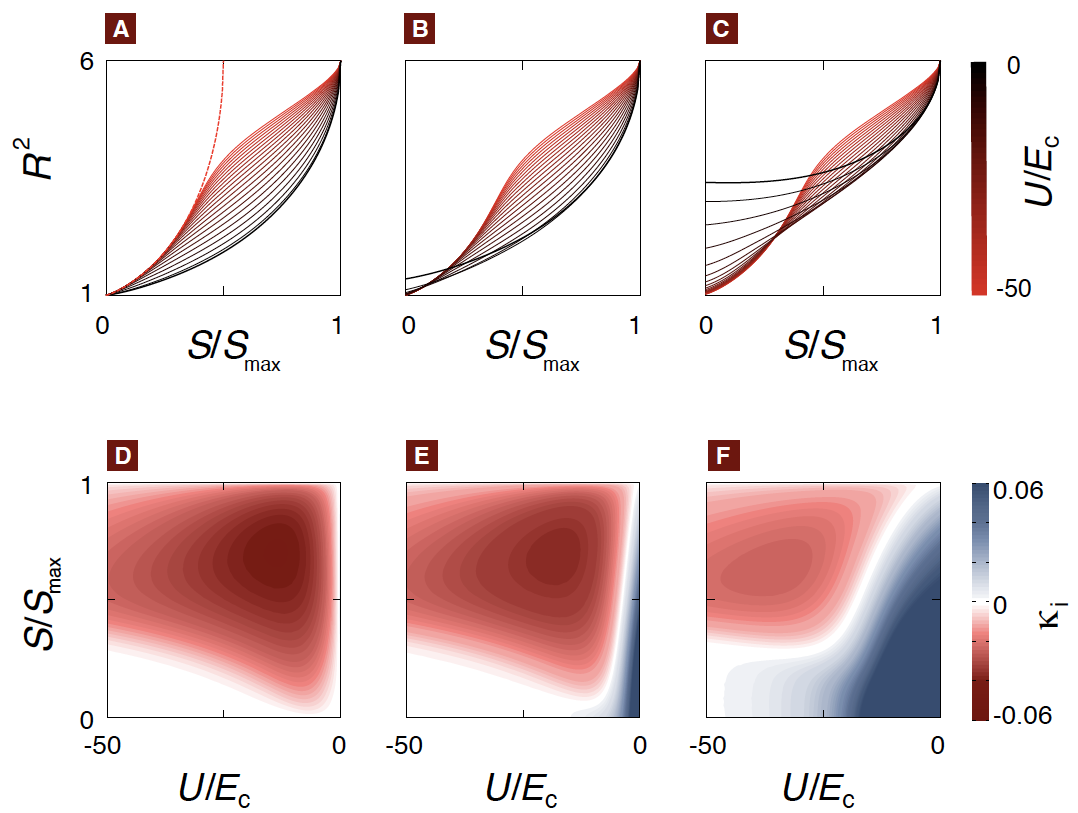}
\end{center}
\caption{{\bf Transition from positive to negative compressibility at finite tunneling}. Exact diagonalization of a one-dimensional system with $6$ particles and $7$ sites.
	{\bf A-C}, squared radius versus entropy for increasingly attractive interactions (black to red curves) at different values of the tunneling amplitude, $t/E_c=0$ ({\bf A}), $t/E_c=2.25$ ({\bf B}) and $t/E_c=6.75$ ({\bf C}). Entropy is given in units of the maximum entropy of the system, $S_{\text{max}}$. The red dashed line in ({\bf A}) corresponds to the limiting case of $U=-\infty$.
For any non-vanishing tunneling the curves eventually cross as the attraction increases, indicating a change in the sign of the compressibility. This crossing is shifted to larger values of attraction and entropy as tunneling increases. {\bf D}, {\bf E}, {\bf F}, interaction compressibility versus entropy and interaction strength at different values of tunneling (same as for {\bf A}, {\bf B}, {\bf C}). Regions of negative (positive) compressibility are marked in red (blue) color. The white color highlights the zero compressibility region where the minimum size of the system is reached.\label{fig:EntropyRadius}}
\end{figure}

The value of attractive interaction above which the compressibility becomes negative should increase with tunneling. As tunneling gets larger the role of interactions is effectively diminished and a larger interaction is required for the entropic effect to overcome the energy minimization effect.

The above predictions can be illustrated by exact diagonalization of a small system (see Figs.~\ref{fig:RadiusCompetition},~\ref{fig:EntropyRadius}).
The full many-body problem cannot be solved exactly, due to the strong correlations involved. In order to analyze a three-dimensional many-particle system we have used a high temperature expansion \cite{Metzner:1991,Henderson:1992,Scarola:2009} (see appendix). This approximation treats interaction exactly and applies when tunneling is much smaller than either interaction or temperature. Already the first two terms of the high temperature expansion capture the competition between energy and entropy and predict the nontrivial minimum in the cloud radius (see Fig. ~\ref{fig:Comparison}B).

\subsection{Comparison between theory and experiment}
\begin{figure}
\begin{center}
		\includegraphics[width=\columnwidth]{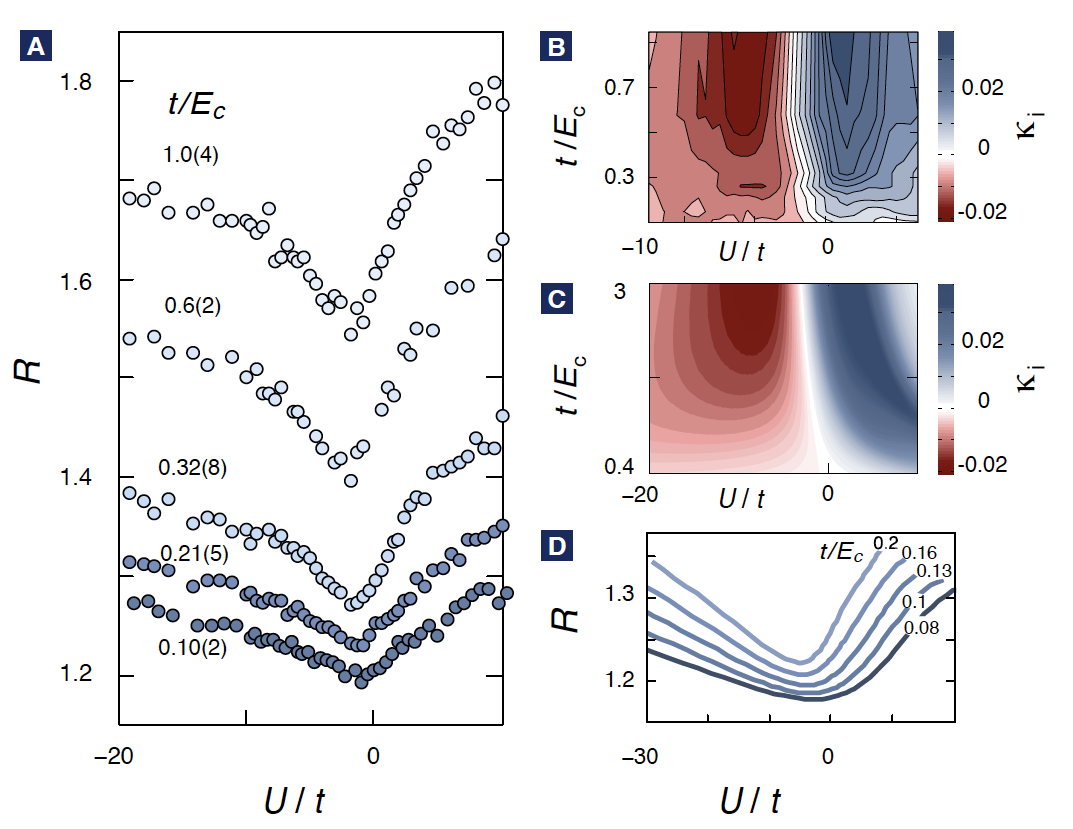}
\end{center}	
\caption{{\bf Comparison of experimental and theoretical size behaviour.}
	{\bf A}, measured rescaled radius $R$ versus interaction strength for different external confinements. Lattice depth is fixed to $7E_r$ and entropy is $S/(k_BN)=0.9-1.4$ ($T/T_F=0.12(3)$). The radius is rescaled in units of the radius of a maximally packed system as defined in the text. For each  data set the corresponding ratio between tunneling and confinement energy $t/E_c$ is indicated. 
{\bf B}, experimental and {\bf C}, calculated compressibility via exact diagonalization of a small system at $S/(k_BN)=0.7$. Compressibilities are plotted versus interaction strength as a function of $t/E_c$.
Experimental compressibilities are determined from the measured cloud size via a linear fit to ten consecutive data points. 
{\bf D} Calculated radius of an atom cloud in a 3D lattice using a high temperature expansion for different external confinements at fixed entropy $S/(k_BN)=1.6$.\label{fig:Comparison}}	
\end{figure}

In Fig.~\ref{fig:Comparison}A we show the experimental data obtained at fixed lattice depth for different external confinements, for which the ratios $U/t$ and $t/E_c$ are varied independently.
The experimental observation and the theoretical prediction show the same qualitative features. As predicted above, for increasing tunneling (decreasing confinement) the observed transition from positive to negative compressibility shifts to stronger attractive interactions (Figs.\,\ref{fig:Comparison}B,C). It is interesting to note that for large ratios of $t/E_c$, where the role of the external confinement becomes unimportant, the transition occurs at a nearly constant value of $U/t$, the only energy scale left in the problem. 

The size expansion of the gas observed in the experiment when increasing the interaction from zero to the maximum experimental negative value ($\vert U/t \vert \sim 20$) is on the order of $5-8\%$. This is in agreement with the expansion we would expect assuming that the non-interacting gas is adiabatically converted into a gas of spinless fermions ($\sim 6-11\%$). Moreover, in order to rule out a possible size increase due to heating, temperatures after reversing the lattice loading process have been measured for all scattering lengths (see appendix). For large values of $t/E_c$, where the effect is most pronounced, a very small heating is observed ($\sim 1\%$ of $T_F$), which could only account for a negligible expansion of the gas (up to $\sim 1\%$), below the experimental shot to shot variation.

In the low temperature and large tunneling regime for which the anomalous expansion effect is most pronounced in the experiment, the high temperature expansion cannot be applied anymore. For lower tunneling and stronger confinement the experiment is affected by larger heating rates, which could mask the isentropic expansion, making a full quantitative comparison difficult. We note, however, that on a qualitative level, all trends such as the steepening of the anomalous expansion with increasing $t/E_c$ agree very well between experiment and theory.

\subsection{Conclusions and outlook}
We have demonstrated how pair formation in an attractively interacting spin mixture of fermionic atoms in an optical lattice gives rise to an anomalous expansion of the gas as the attraction increases. This novel effect is the result of the quenching of quantum fluctuations caused by pairing in a lattice potential, which effectively enhances the role of entropy and fermionizes the bosonic pairs. 

Our observation reveals for the first time the fundamentally different consequences of pairing in the first band of a lattice potential as compared to those in the continuum. It constitutes a step towards the study of superfluidity in the fermionic Hubbard model, where characterization and detection of the many-body paired states, especially the pseudogap regime relevant to high temperature superconductivity, is a major goal.

Examples of exotic thermodynamic behaviour caused by the interplay of strong interactions and entropy have been scarcely observed in quantum many-body systems. Our work might pave the way towards the discovery of other novel instances of this type of phenomena with cold atomic gases. It remains an important experimental and theoretical challenge to identify such novel phenomena and design appropriate protocols to observe them in the laboratory. 
We anticipate that similar effects can occur for attractive Fermi mixtures with population imbalance \cite{Partridge:2006,Zwierlein:2006}, where the intriguing features of pairing of fermions with different Fermi energies are the focus of current investigation and controversy. 

Our work also opens an interesting route towards the detection of quantum many-body phases at finite entropies, where a dramatic change in the thermodynamic behaviour can serve as a footprint of the crossover between two phases exhibiting substantially different entropy densities, as observed recently for a quantum critical system \cite{Rost:2009}. This might be a promising perspective for the detection of transitions between two  topological phases \cite{Nayak:2008}, whose different topology can lead to strikingly distinctive ways of storing entropy \cite{Cooper:2009}.



\begin{acknowledgements}
We would like to thank W.\,Zwerger for stimulating discussions. This work was supported by the DFG, the EU (IP SCALA), EuroQUAM (LH), DARPA (OLE program), AFOSR, DIP (EA and IB), US Israel BSF (EA and ED) and MATCOR (SW).   

Correspondence and requests for materials should be addressed to B.\,Paredes~(email: paredes@uni-mainz.de).
\end{acknowledgements}

\appendix 
\section{Supplementary Information}	
\subsection{Thermodynamics at zero tunneling}
At zero tunneling the partition function $Z_0$ of the system is a product of local partition functions at each site $Z_0=\prod_\ell z_\ell$ with
\begin{equation}
z_\ell=1+2e^{-\beta(E_cr_\ell^2-\mu-\frac{U}{2})}+e^{-2\beta(E_cr_\ell^2-\mu)},
\end{equation}
where $\beta=1/k_BT$ and $T$ is the temperature of the gas.
Defining the probabilities $p_\ell^0=1/z_\ell$, $p_\ell^{1}=e^{-\beta(E_c r_\ell^2-\mu-\frac{U}{2})}/z_\ell$, $p_\ell^{2}=e^{-2\beta(E_c r_\ell^2-\mu)}/z_\ell$ for zero, single and double site occupation, respectively, the local occupation per spin state is given by $n_{\ell}= p_\ell^{1}+p_\ell^{2}$ and the entropy per site is $s_\ell=-p_\ell^{0} \log p_\ell^{0} - 2 p_\ell^{1} \log p_\ell^{1} - p_\ell^{2} \log p_\ell^{2}$. For a given entropy $S$ and a number of particles per spin state $N_{\sigma}$ the radius of the system $R$ at any  interaction strength $U$ can be calculated by imposing the conditions $\sum_\ell s_\ell=S$ and $\sum_\ell n_\ell=N_{\sigma}$.


\subsection{Entropy density reduction and size expansion at zero tunneling}
At zero tunneling in the two limiting cases of $U=0$ and $U=-\infty$ the system behaves as a gas of non-interacting fermions with local Hamiltonian
\begin{equation}
	\hat{h}_\ell \propto r_\ell^2 \hat{n}_\ell.
	\label{local}
\end{equation}
For $U=0$ it is a two-component gas, whereas for $U=-\infty$ it behaves as a single component made out of fermions with twice the mass.  The thermodynamic magnitudes for both cases can be directly obtained by relating them to those of a single component with the same local Hamiltonian.
Let us denote by $z_\ell^\prime(\beta)$, $n_\ell^\prime(\beta)$ and $s_\ell^\prime(\beta)$ the local partition function, density and entropy, respectively, of a single component gas at temperature $k_BT=1/\beta$. 
For a two-component gas, consisting of two copies of that one, the corresponding thermodynamic magnitudes are $\left(z_\ell^\prime (\beta)\right)^2$,  $n_\ell^\prime(\beta)$, and $2s_\ell^\prime(\beta)$.
In contrast, for a single component with double mass, energies are multiplied by a factor $2$ and the corresponding thermodynamic functions are  $z_\ell^\prime(2\beta)$, $n_\ell^\prime(2\beta)$ and $s_\ell^\prime(2\beta)$.
The one-component system with double mass has therefore the same radius (same density per component) as the two-component gas when its temperature is twice as high. In that situation its entropy is exactly half the one of the two-component gas.

The relative expansion of the gas as the interaction is adiabatically increased from $U=0$ to $U=-\infty$ can be directly estimated for low entropies. 
As for a free gas, for temperatures much smaller than the Fermi temperature the entropy changes linearly with temperature, while the energy increase is quadratic. For our system, with  $E\propto R^2$, the squared radius increases quadratically with entropy (see Fig.~\ref{fig:ZeroTunneling}A), $R^2 \sim 1+ \alpha (S/N)^2$, where $\alpha$ can be found to be $k_BT_F/\pi^2E_c=5/3\pi^2$. Since the number of particles is half for the one-component gas we have that the relative radius increase is $\Delta R (U=-\infty) / \Delta R (U=0) \sim 2$. 

\subsection{High Temperature Expansion}
The so-called high temperature expansion is a useful analytical approach to analyze the Hubbard model. Within this method the partition function of the system $Z=$tr$(e^{-\beta \hat{H}})$ is expanded around the one at zero tunneling $Z_0$, which can be calculated exactly as shown above. The logarithm of the partition function can be formally written as:
\begin{equation}
\begin{aligned}
\log(Z) = & \log(Z_{0}) + \\  
          & \log \left\langle \textrm{T}\left\{ \exp \left( - \int^{\beta}_{0} \hat{K} (\tau' ) d\tau' \right) \right\} \right\rangle_{0},
\label{expansion}
\end{aligned}
\end{equation}

where $\hat{K}=-t\sum_{\langle \ell , \ell'\rangle \sigma}
 c^{\dagger}_{\ell\sigma}
c^{\,}_{\ell'\sigma}$ is the tunneling Hamiltonian, $\textrm{T}$ denotes imaginary time ordering and $\langle \rangle_{0}$ is the expectation value calculated with the unperturbed partition function $Z_0$. Expression (\ref{expansion}) can be used to perform a systematic expansion of $Z$ in  powers of $\beta t$. The radius can be then obtained as $R^2=- 1 / (\beta N_\sigma) \partial (\log Z)/ \partial E_c \vert_{\beta, \mu}$.
This method treats interaction and confinement exactly and is valid for values of tunneling much smaller than temperature, interaction and confinement energy. Already a second order expansion can efficiently describe the competition between the Hartree interaction, which induces the compression of the cloud for weak interactions, and entropy, which gives rise to the expansion of the gas for strong attraction.

\subsection{Image analysis}
The phase contrast images were fitted using two-dimensional adapted Fermi-Dirac fits
\begin{equation}
\begin{aligned}
F(x,y)=&a\, \text{Li}_2\left(-100\,e^{-\frac{(x-x_c)^2}{2\sigma_x^2}-\frac{(y-y_c)^2}{2\sigma_y^2}}\right)+ \\   
        &b+c\,\sqrt{\frac{(x-x_c)^2}{\sigma_x^2}+\frac{(y-y_c)^2}{\sigma_y^2}}, 
\end{aligned} 
\end{equation}

with $\text{Li}_2$ being the di-logarithm and $x_c,y_c,\sigma_x,\sigma_y,a,b,c$ free fit parameters.
The perpendicular cloud size $R_\bot$ after integration over the propagation axis of the imaging laser beam was extracted from the fits through $R_\bot=\sqrt{1.264^2\, (\sigma_x^2+\sigma_y^2)-w^2}$, where $w$ denotes the imaging resolution (Radius of Airy disc $< 3\mu$m) of our imaging setup.
The renormalized radius $R$ is then given by $R=\sqrt{3/2}R_\bot/r_c$.

\subsection{Lattice Calibration and Ramps}
The lattice depth was calibrated using frequency modulation of the lattice beams to resonantly excite atomic population to higher vibrational energy bands. The lattice was ramped to the final depth using linear ramps. The ramp rates  were chosen such that a minimum cloud size was obtained for low confinement. For much shorter ramp times non-adiabatic effects lead to a larger cloud size, whereas for much longer ramp times heating leads to increased cloud sizes. For the case of strong attractive interactions and strong confinements $(t/E_c<0.2)$, we find that the cloud size has increased beyond the minimum size again for the ramp times used. We attribute this behaviour to increased heating observed for strong attractive interactions.

\subsection{Temperature Measurements and Heating Rates}
All quoted temperatures were determined from Fermi-Dirac fits to time-of-flight absorption images of a non-interacting spin mixture, released from the dipole trap at low compression ($\omega_\perp\simeq 2\pi \times 25$\,Hz). For temperatures below $0.15 T_F$ we note that the reliability of the fit deteriorates, resulting in an increased temperature uncertainty. The non-interacting cloud size can serve as an additional thermometer and would suggest initial temperatures around $0.1 T_F$.

We have measured the entropy increase due to losses and technical noise in the lattice by loading and unloading the atoms from the lattice and determining the increase in $T/T_F$, and assuming the same heating during loading and unloading. For low confinements $(t/E_c>0.5)$ the measured maximum temperature increase per ramp is $<0.02 T_F$. At stronger confinements ($t/E_c<0.5$) the heating increases with increasing density and increasing attractive interaction. In the interaction range, where we observe the cloud size minimum $(-5<U/t<0)$, the temperature increase varies between $0.01 T_F$ for $U/t=0$ and $0.02 T_F$ for $U/t=-5$. For the strongest interactions and strong confinement $(t/E_c<0.2)$ we observe heatings up to $0.06 T_F$. The main temperature increase may be caused by pair losses, inelastic three-body collisions and technical noise.


\bibliographystyle{apsrev_woURL}
\bibliography{References_NegativeU}

\begin{thebibliography}{44}
\expandafter\ifx\csname natexlab\endcsname\relax\def\natexlab#1{#1}\fi
\expandafter\ifx\csname bibnamefont\endcsname\relax
  \def\bibnamefont#1{#1}\fi
\expandafter\ifx\csname bibfnamefont\endcsname\relax
  \def\bibfnamefont#1{#1}\fi
\expandafter\ifx\csname citenamefont\endcsname\relax
  \def\citenamefont#1{#1}\fi
\expandafter\ifx\csname url\endcsname\relax
  \def\url#1{\texttt{#1}}\fi
\expandafter\ifx\csname urlprefix\endcsname\relax\def\urlprefix{URL }\fi
\providecommand{\bibinfo}[2]{#2}
\providecommand{\eprint}[2][]{\url{#2}}

\bibitem[{\citenamefont{Auerbach}(2006)}]{Auerbachbook}
\bibinfo{author}{\bibfnamefont{A.}~\bibnamefont{Auerbach}},
  \emph{\bibinfo{title}{Interacting Electrons and Quantum Magnetism}}
  (\bibinfo{publisher}{Springer}, \bibinfo{year}{2006}).

\bibitem[{\citenamefont{Wen}(2004)}]{Wen:2004}
\bibinfo{author}{\bibfnamefont{X.~G.} \bibnamefont{Wen}},
  \emph{\bibinfo{title}{Quantum Field Theory of Many-Body Systems}}, Oxford
  Graduate Texts (\bibinfo{publisher}{Oxford University Press},
  \bibinfo{address}{Oxford}, \bibinfo{year}{2004}).

\bibitem[{\citenamefont{Pomeranchuk}(1950)}]{Pomeranchuk:1950}
\bibinfo{author}{\bibfnamefont{I.}~\bibnamefont{Pomeranchuk}},
  \bibinfo{journal}{Zh. Eksp. Teor. Fiz.} \textbf{\bibinfo{volume}{20}},
  \bibinfo{pages}{919} (\bibinfo{year}{1950}).

\bibitem[{\citenamefont{Richardson}(1997)}]{Richardson:1997}
\bibinfo{author}{\bibfnamefont{R.~C.} \bibnamefont{Richardson}},
  \bibinfo{journal}{Rev. Mod. Phys.} \textbf{\bibinfo{volume}{69}},
  \bibinfo{pages}{683} (\bibinfo{year}{1997}).

\bibitem[{\citenamefont{Jaksch and Zoller}(2005)}]{Jaksch:2005}
\bibinfo{author}{\bibfnamefont{D.}~\bibnamefont{Jaksch}} \bibnamefont{and}
  \bibinfo{author}{\bibfnamefont{P.}~\bibnamefont{Zoller}},
  \bibinfo{journal}{Ann. Phys. (N.Y.)} \textbf{\bibinfo{volume}{315}},
  \bibinfo{pages}{52} (\bibinfo{year}{2005}).

\bibitem[{\citenamefont{Lewenstein et~al.}(2007)\citenamefont{Lewenstein,
  Sanpera, Ahufinger, Damski, De, and Sen}}]{Lewenstein:2007}
\bibinfo{author}{\bibfnamefont{M.}~\bibnamefont{Lewenstein}},
  \bibinfo{author}{\bibfnamefont{A.}~\bibnamefont{Sanpera}},
  \bibinfo{author}{\bibfnamefont{V.}~\bibnamefont{Ahufinger}},
  \bibinfo{author}{\bibfnamefont{B.}~\bibnamefont{Damski}},
  \bibinfo{author}{\bibfnamefont{A.~S.} \bibnamefont{De}}, \bibnamefont{and}
  \bibinfo{author}{\bibfnamefont{U.}~\bibnamefont{Sen}}, \bibinfo{journal}{Adv.
  Phys.} \textbf{\bibinfo{volume}{56}}, \bibinfo{pages}{243}
  (\bibinfo{year}{2007}).

\bibitem[{\citenamefont{Bloch et~al.}(2008)\citenamefont{Bloch, Dalibard, and
  Zwerger}}]{Bloch:2008c}
\bibinfo{author}{\bibfnamefont{I.}~\bibnamefont{Bloch}},
  \bibinfo{author}{\bibfnamefont{J.}~\bibnamefont{Dalibard}}, \bibnamefont{and}
  \bibinfo{author}{\bibfnamefont{W.}~\bibnamefont{Zwerger}},
  \bibinfo{journal}{Rev. Mod. Phys.} \textbf{\bibinfo{volume}{80}},
  \bibinfo{eid}{885} (pages~\bibinfo{numpages}{80}) (\bibinfo{year}{2008}).

\bibitem[{\citenamefont{Fisher et~al.}(1989)\citenamefont{Fisher, Weichman,
  Grinstein, and Fisher}}]{Fisher:1989}
\bibinfo{author}{\bibfnamefont{M.~P.~A.} \bibnamefont{Fisher}},
  \bibinfo{author}{\bibfnamefont{P.~B.} \bibnamefont{Weichman}},
  \bibinfo{author}{\bibfnamefont{G.}~\bibnamefont{Grinstein}},
  \bibnamefont{and} \bibinfo{author}{\bibfnamefont{D.~S.}
  \bibnamefont{Fisher}}, \bibinfo{journal}{Phys. Rev. B}
  \textbf{\bibinfo{volume}{40}}, \bibinfo{pages}{546} (\bibinfo{year}{1989}).

\bibitem[{\citenamefont{Jaksch et~al.}(1998)\citenamefont{Jaksch, Bruder,
  Cirac, Gardiner, and Zoller}}]{Jaksch:1998}
\bibinfo{author}{\bibfnamefont{D.}~\bibnamefont{Jaksch}},
  \bibinfo{author}{\bibfnamefont{C.}~\bibnamefont{Bruder}},
  \bibinfo{author}{\bibfnamefont{J.~I.} \bibnamefont{Cirac}},
  \bibinfo{author}{\bibfnamefont{C.~W.} \bibnamefont{Gardiner}},
  \bibnamefont{and} \bibinfo{author}{\bibfnamefont{P.}~\bibnamefont{Zoller}},
  \bibinfo{journal}{Phys. Rev. Lett} \textbf{\bibinfo{volume}{81}},
  \bibinfo{pages}{3108} (\bibinfo{year}{1998}).

\bibitem[{\citenamefont{Greiner et~al.}(2002)\citenamefont{Greiner, Mandel,
  Esslinger, H{\"a}nsch, and Bloch}}]{Greiner:2002a}
\bibinfo{author}{\bibfnamefont{M.}~\bibnamefont{Greiner}},
  \bibinfo{author}{\bibfnamefont{M.~O.} \bibnamefont{Mandel}},
  \bibinfo{author}{\bibfnamefont{T.}~\bibnamefont{Esslinger}},
  \bibinfo{author}{\bibfnamefont{T.~W.} \bibnamefont{H{\"a}nsch}},
  \bibnamefont{and} \bibinfo{author}{\bibfnamefont{I.}~\bibnamefont{Bloch}},
  \bibinfo{journal}{Nature} \textbf{\bibinfo{volume}{415}}, \bibinfo{pages}{39}
  (\bibinfo{year}{2002}).

\bibitem[{\citenamefont{Paredes et~al.}(2004)\citenamefont{Paredes, Widera,
  Murg, Mandel, {F\"{o}lling}, Cirac, Shlyapnikov, {H\"{a}nsch}, and
  Bloch}}]{Paredes:2004}
\bibinfo{author}{\bibfnamefont{B.}~\bibnamefont{Paredes}},
  \bibinfo{author}{\bibfnamefont{A.}~\bibnamefont{Widera}},
  \bibinfo{author}{\bibfnamefont{V.}~\bibnamefont{Murg}},
  \bibinfo{author}{\bibfnamefont{O.}~\bibnamefont{Mandel}},
  \bibinfo{author}{\bibfnamefont{S.}~\bibnamefont{{F\"{o}lling}}},
  \bibinfo{author}{\bibfnamefont{J.~I.} \bibnamefont{Cirac}},
  \bibinfo{author}{\bibfnamefont{G.~V.} \bibnamefont{Shlyapnikov}},
  \bibinfo{author}{\bibfnamefont{T.~W.} \bibnamefont{{H\"{a}nsch}}},
  \bibnamefont{and} \bibinfo{author}{\bibfnamefont{I.}~\bibnamefont{Bloch}},
  \bibinfo{journal}{Nature} \textbf{\bibinfo{volume}{429}}, \bibinfo{pages}{277
  } (\bibinfo{year}{2004}).

\bibitem[{\citenamefont{Kinoshita et~al.}(2004)\citenamefont{Kinoshita, Wenger,
  and Weiss}}]{Kinoshita:2004}
\bibinfo{author}{\bibfnamefont{T.}~\bibnamefont{Kinoshita}},
  \bibinfo{author}{\bibfnamefont{T.}~\bibnamefont{Wenger}}, \bibnamefont{and}
  \bibinfo{author}{\bibfnamefont{D.~S.} \bibnamefont{Weiss}},
  \bibinfo{journal}{Science} \textbf{\bibinfo{volume}{305}},
  \bibinfo{pages}{1125} (\bibinfo{year}{2004}).

\bibitem[{\citenamefont{St{\"o}ferle et~al.}(2004)\citenamefont{St{\"o}ferle,
  Moritz, Schori, K{\"o}hl, and Esslinger}}]{Stoferle:2004}
\bibinfo{author}{\bibfnamefont{T.}~\bibnamefont{St{\"o}ferle}},
  \bibinfo{author}{\bibfnamefont{H.}~\bibnamefont{Moritz}},
  \bibinfo{author}{\bibfnamefont{C.}~\bibnamefont{Schori}},
  \bibinfo{author}{\bibfnamefont{M.}~\bibnamefont{K{\"o}hl}}, \bibnamefont{and}
  \bibinfo{author}{\bibfnamefont{T.}~\bibnamefont{Esslinger}},
  \bibinfo{journal}{Phys. Phys. Lett.} \textbf{\bibinfo{volume}{92}},
  \bibinfo{pages}{130403} (\bibinfo{year}{2004}).

\bibitem[{\citenamefont{Spielman et~al.}(2007)\citenamefont{Spielman, Phillips,
  and Porto}}]{Spielman:2007}
\bibinfo{author}{\bibfnamefont{I.~B.} \bibnamefont{Spielman}},
  \bibinfo{author}{\bibfnamefont{W.~D.} \bibnamefont{Phillips}},
  \bibnamefont{and} \bibinfo{author}{\bibfnamefont{J.~V.} \bibnamefont{Porto}},
  \bibinfo{journal}{Phys. Rev. Lett.} \textbf{\bibinfo{volume}{98}},
  \bibinfo{eid}{080404} (pages~\bibinfo{numpages}{4}) (\bibinfo{year}{2007}).

\bibitem[{\citenamefont{Mun et~al.}(2007)\citenamefont{Mun, Medley, Campbell,
  Marcassa, Pritchard, and Ketterle}}]{Mun:2007}
\bibinfo{author}{\bibfnamefont{J.}~\bibnamefont{Mun}},
  \bibinfo{author}{\bibfnamefont{P.}~\bibnamefont{Medley}},
  \bibinfo{author}{\bibfnamefont{G.~K.} \bibnamefont{Campbell}},
  \bibinfo{author}{\bibfnamefont{L.~G.} \bibnamefont{Marcassa}},
  \bibinfo{author}{\bibfnamefont{D.~E.} \bibnamefont{Pritchard}},
  \bibnamefont{and} \bibinfo{author}{\bibfnamefont{W.}~\bibnamefont{Ketterle}},
  \bibinfo{journal}{Phys. Rev. Lett.} \textbf{\bibinfo{volume}{99}},
  \bibinfo{eid}{150604} (pages~\bibinfo{numpages}{4}) (\bibinfo{year}{2007}).

\bibitem[{\citenamefont{Fallani et~al.}(2007)\citenamefont{Fallani, Lye,
  Guarrera, Fort, and Inguscio}}]{Fallani:2007}
\bibinfo{author}{\bibfnamefont{L.}~\bibnamefont{Fallani}},
  \bibinfo{author}{\bibfnamefont{J.~E.} \bibnamefont{Lye}},
  \bibinfo{author}{\bibfnamefont{V.}~\bibnamefont{Guarrera}},
  \bibinfo{author}{\bibfnamefont{C.}~\bibnamefont{Fort}}, \bibnamefont{and}
  \bibinfo{author}{\bibfnamefont{M.}~\bibnamefont{Inguscio}},
  \bibinfo{journal}{Phys. Rev. Lett.} \textbf{\bibinfo{volume}{98}},
  \bibinfo{eid}{130404} (pages~\bibinfo{numpages}{4}) (\bibinfo{year}{2007}).

\bibitem[{\citenamefont{Spielman et~al.}(2008)\citenamefont{Spielman, Phillips,
  and Porto}}]{Spielman:2008}
\bibinfo{author}{\bibfnamefont{I.~B.} \bibnamefont{Spielman}},
  \bibinfo{author}{\bibfnamefont{W.~D.} \bibnamefont{Phillips}},
  \bibnamefont{and} \bibinfo{author}{\bibfnamefont{J.~V.} \bibnamefont{Porto}},
  \bibinfo{journal}{Phys. Rev. Lett.} \textbf{\bibinfo{volume}{100}},
  \bibinfo{eid}{120402} (pages~\bibinfo{numpages}{4}) (\bibinfo{year}{2008}).

\bibitem[{\citenamefont{Guarrera et~al.}(2008)\citenamefont{Guarrera, Fabbri,
  Fallani, Fort, van~der Stam, and Inguscio}}]{Guarrera:2008}
\bibinfo{author}{\bibfnamefont{V.}~\bibnamefont{Guarrera}},
  \bibinfo{author}{\bibfnamefont{N.}~\bibnamefont{Fabbri}},
  \bibinfo{author}{\bibfnamefont{L.}~\bibnamefont{Fallani}},
  \bibinfo{author}{\bibfnamefont{C.}~\bibnamefont{Fort}},
  \bibinfo{author}{\bibfnamefont{K.~M.~R.} \bibnamefont{van~der Stam}},
  \bibnamefont{and} \bibinfo{author}{\bibfnamefont{M.}~\bibnamefont{Inguscio}},
  \bibinfo{journal}{Phys. Rev. Lett.} \textbf{\bibinfo{volume}{100}},
  \bibinfo{eid}{250403} (pages~\bibinfo{numpages}{4}) (\bibinfo{year}{2008}).

\bibitem[{\citenamefont{Chin et~al.}(2006)\citenamefont{Chin, Miller, Liu,
  Stan, Setiawan, Sanner, Xu, and Ketterle}}]{Chin:2006}
\bibinfo{author}{\bibfnamefont{J.}~\bibnamefont{Chin}},
  \bibinfo{author}{\bibfnamefont{D.}~\bibnamefont{Miller}},
  \bibinfo{author}{\bibfnamefont{Y.}~\bibnamefont{Liu}},
  \bibinfo{author}{\bibfnamefont{C.}~\bibnamefont{Stan}},
  \bibinfo{author}{\bibfnamefont{W.}~\bibnamefont{Setiawan}},
  \bibinfo{author}{\bibfnamefont{C.}~\bibnamefont{Sanner}},
  \bibinfo{author}{\bibfnamefont{K.}~\bibnamefont{Xu}}, \bibnamefont{and}
  \bibinfo{author}{\bibfnamefont{W.}~\bibnamefont{Ketterle}},
  \bibinfo{journal}{Nature} \textbf{\bibinfo{volume}{443}},
  \bibinfo{pages}{961} (\bibinfo{year}{2006}).

\bibitem[{\citenamefont{Strohmaier et~al.}(2007)\citenamefont{Strohmaier,
  Takasu, G\"{u}nter, J\"{o}rdens, K\"{o}hl, Moritz, and
  Esslinger}}]{Strohmaier:2007}
\bibinfo{author}{\bibfnamefont{N.}~\bibnamefont{Strohmaier}},
  \bibinfo{author}{\bibfnamefont{Y.}~\bibnamefont{Takasu}},
  \bibinfo{author}{\bibfnamefont{K.}~\bibnamefont{G\"{u}nter}},
  \bibinfo{author}{\bibfnamefont{R.}~\bibnamefont{J\"{o}rdens}},
  \bibinfo{author}{\bibfnamefont{M.}~\bibnamefont{K\"{o}hl}},
  \bibinfo{author}{\bibfnamefont{H.}~\bibnamefont{Moritz}}, \bibnamefont{and}
  \bibinfo{author}{\bibfnamefont{T.}~\bibnamefont{Esslinger}},
  \bibinfo{journal}{Phys. Rev. Lett.} \textbf{\bibinfo{volume}{99}},
  \bibinfo{eid}{220601} (pages~\bibinfo{numpages}{4}) (\bibinfo{year}{2007}).

\bibitem[{\citenamefont{J\"ordens et~al.}(2008)\citenamefont{J\"ordens,
  Strohmaier, G\"unter, Moritz, and Esslinger}}]{Jordens:2008}
\bibinfo{author}{\bibfnamefont{R.}~\bibnamefont{J\"ordens}},
  \bibinfo{author}{\bibfnamefont{N.}~\bibnamefont{Strohmaier}},
  \bibinfo{author}{\bibfnamefont{K.}~\bibnamefont{G\"unter}},
  \bibinfo{author}{\bibfnamefont{H.}~\bibnamefont{Moritz}}, \bibnamefont{and}
  \bibinfo{author}{\bibfnamefont{T.}~\bibnamefont{Esslinger}},
  \bibinfo{journal}{Nature} \textbf{\bibinfo{volume}{455}},
  \bibinfo{pages}{204} (\bibinfo{year}{2008}).

\bibitem[{\citenamefont{Schneider et~al.}(2008)\citenamefont{Schneider,
  Hackerm\"uller, Will, Best, Bloch, Costi, Helmes, Rasch, and
  Rosch}}]{Schneider:2008a}
\bibinfo{author}{\bibfnamefont{U.}~\bibnamefont{Schneider}},
  \bibinfo{author}{\bibfnamefont{L.}~\bibnamefont{Hackerm\"uller}},
  \bibinfo{author}{\bibfnamefont{S.}~\bibnamefont{Will}},
  \bibinfo{author}{\bibfnamefont{T.}~\bibnamefont{Best}},
  \bibinfo{author}{\bibfnamefont{I.}~\bibnamefont{Bloch}},
  \bibinfo{author}{\bibfnamefont{T.~A.} \bibnamefont{Costi}},
  \bibinfo{author}{\bibfnamefont{R.~W.} \bibnamefont{Helmes}},
  \bibinfo{author}{\bibfnamefont{D.}~\bibnamefont{Rasch}}, \bibnamefont{and}
  \bibinfo{author}{\bibfnamefont{A.}~\bibnamefont{Rosch}},
  \bibinfo{journal}{Science} \textbf{\bibinfo{volume}{322}},
  \bibinfo{pages}{1520} (\bibinfo{year}{2008}).

\bibitem[{\citenamefont{Regal et~al.}(2004)\citenamefont{Regal, Greiner, and
  Jin}}]{Regal:2004a}
\bibinfo{author}{\bibfnamefont{C.~A.} \bibnamefont{Regal}},
  \bibinfo{author}{\bibfnamefont{M.}~\bibnamefont{Greiner}}, \bibnamefont{and}
  \bibinfo{author}{\bibfnamefont{D.~S.} \bibnamefont{Jin}},
  \bibinfo{journal}{Phys. Rev. Lett.} \textbf{\bibinfo{volume}{92}},
  \bibinfo{pages}{040403} (\bibinfo{year}{2004}).

\bibitem[{\citenamefont{Zwierlein et~al.}(2004)\citenamefont{Zwierlein, Stan,
  Schunck, Raupach, Kerman, and Ketterle}}]{Zwierlein:2004}
\bibinfo{author}{\bibfnamefont{M.~W.} \bibnamefont{Zwierlein}},
  \bibinfo{author}{\bibfnamefont{C.~A.} \bibnamefont{Stan}},
  \bibinfo{author}{\bibfnamefont{C.~H.} \bibnamefont{Schunck}},
  \bibinfo{author}{\bibfnamefont{S.~M.~F.} \bibnamefont{Raupach}},
  \bibinfo{author}{\bibfnamefont{A.~J.} \bibnamefont{Kerman}},
  \bibnamefont{and} \bibinfo{author}{\bibfnamefont{W.}~\bibnamefont{Ketterle}},
  \bibinfo{journal}{Phys. Rev. Lett.} \textbf{\bibinfo{volume}{92}},
  \bibinfo{pages}{120403} (\bibinfo{year}{2004}).

\bibitem[{\citenamefont{Bartenstein et~al.}(2004)\citenamefont{Bartenstein,
  Altmeyer, Riedl, Joachim, Chin, Hecker-Denschlag, and
  Grimm}}]{Bartenstein:2004b}
\bibinfo{author}{\bibfnamefont{M.}~\bibnamefont{Bartenstein}},
  \bibinfo{author}{\bibfnamefont{A.}~\bibnamefont{Altmeyer}},
  \bibinfo{author}{\bibfnamefont{S.}~\bibnamefont{Riedl}},
  \bibinfo{author}{\bibfnamefont{S.}~\bibnamefont{Joachim}},
  \bibinfo{author}{\bibfnamefont{C.}~\bibnamefont{Chin}},
  \bibinfo{author}{\bibfnamefont{J.}~\bibnamefont{Hecker-Denschlag}},
  \bibnamefont{and} \bibinfo{author}{\bibfnamefont{R.}~\bibnamefont{Grimm}},
  \bibinfo{journal}{Phys. Rev. Lett.} \textbf{\bibinfo{volume}{92}},
  \bibinfo{pages}{120401} (\bibinfo{year}{2004}).

\bibitem[{\citenamefont{Bourdel et~al.}(2004)\citenamefont{Bourdel, Khaykovich,
  Cubizolles, Zhang, Chevy, Teichmann, Tarruell, S.~J.~H, and
  Salomon}}]{Bourdel:2004}
\bibinfo{author}{\bibfnamefont{T.}~\bibnamefont{Bourdel}},
  \bibinfo{author}{\bibfnamefont{L.}~\bibnamefont{Khaykovich}},
  \bibinfo{author}{\bibfnamefont{M.~E.} \bibnamefont{Cubizolles}},
  \bibinfo{author}{\bibfnamefont{J.}~\bibnamefont{Zhang}},
  \bibinfo{author}{\bibfnamefont{F.}~\bibnamefont{Chevy}},
  \bibinfo{author}{\bibfnamefont{M.}~\bibnamefont{Teichmann}},
  \bibinfo{author}{\bibfnamefont{L.}~\bibnamefont{Tarruell}},
  \bibinfo{author}{\bibfnamefont{M.~F.~K.} \bibnamefont{S.~J.~H}},
  \bibnamefont{and} \bibinfo{author}{\bibfnamefont{C.}~\bibnamefont{Salomon}},
  \bibinfo{journal}{Phys. Rev. Lett.} \textbf{\bibinfo{volume}{93}},
  \bibinfo{pages}{050401} (\bibinfo{year}{2004}).

\bibitem[{\citenamefont{Lee et~al.}(2006)\citenamefont{Lee, Nagaosa, and
  Wen}}]{Lee:2006}
\bibinfo{author}{\bibfnamefont{P.}~\bibnamefont{Lee}},
  \bibinfo{author}{\bibfnamefont{N.}~\bibnamefont{Nagaosa}}, \bibnamefont{and}
  \bibinfo{author}{\bibfnamefont{X.-G.} \bibnamefont{Wen}},
  \bibinfo{journal}{Rev. Mod. Phys.} \textbf{\bibinfo{volume}{78}},
  \bibinfo{pages}{17} (\bibinfo{year}{2006}).

\bibitem[{\citenamefont{Hofstetter et~al.}(2002)\citenamefont{Hofstetter,
  Cirac, Zoller, Demler, and Lukin}}]{Hofstetter:2002}
\bibinfo{author}{\bibfnamefont{W.}~\bibnamefont{Hofstetter}},
  \bibinfo{author}{\bibfnamefont{J.~I.} \bibnamefont{Cirac}},
  \bibinfo{author}{\bibfnamefont{P.}~\bibnamefont{Zoller}},
  \bibinfo{author}{\bibfnamefont{E.}~\bibnamefont{Demler}}, \bibnamefont{and}
  \bibinfo{author}{\bibfnamefont{M.~D.} \bibnamefont{Lukin}},
  \bibinfo{journal}{Phys. Rev. Lett.} \textbf{\bibinfo{volume}{89}},
  \bibinfo{pages}{220407} (\bibinfo{year}{2002}).

\bibitem[{\citenamefont{Toschi et~al.}(2005)\citenamefont{Toschi, Barone,
  Capone, and Castellani}}]{Toschi:2005}
\bibinfo{author}{\bibfnamefont{A.}~\bibnamefont{Toschi}},
  \bibinfo{author}{\bibfnamefont{P.}~\bibnamefont{Barone}},
  \bibinfo{author}{\bibfnamefont{M.}~\bibnamefont{Capone}}, \bibnamefont{and}
  \bibinfo{author}{\bibfnamefont{C.}~\bibnamefont{Castellani}},
  \bibinfo{journal}{New J. Phys.} \textbf{\bibinfo{volume}{7}},
  \bibinfo{pages}{7} (\bibinfo{year}{2005}).

\bibitem[{\citenamefont{Chien et~al.}(2008)\citenamefont{Chien, Chen, and
  Levin}}]{Chien:2008}
\bibinfo{author}{\bibfnamefont{C.-C.} \bibnamefont{Chien}},
  \bibinfo{author}{\bibfnamefont{Q.}~\bibnamefont{Chen}}, \bibnamefont{and}
  \bibinfo{author}{\bibfnamefont{K.}~\bibnamefont{Levin}},
  \bibinfo{journal}{Phys. Rev. A} \textbf{\bibinfo{volume}{78}},
  \bibinfo{eid}{043612} (pages~\bibinfo{numpages}{10}) (\bibinfo{year}{2008}).

\bibitem[{\citenamefont{Paiva et~al.}(2009)\citenamefont{Paiva, Scalettar,
  Randeria, and Trivedi}}]{Paiva:2009}
\bibinfo{author}{\bibfnamefont{T.}~\bibnamefont{Paiva}},
  \bibinfo{author}{\bibfnamefont{R.}~\bibnamefont{Scalettar}},
  \bibinfo{author}{\bibfnamefont{M.}~\bibnamefont{Randeria}}, \bibnamefont{and}
  \bibinfo{author}{\bibfnamefont{N.}~\bibnamefont{Trivedi}},
  \bibinfo{journal}{arXiv:0906.2141}  (\bibinfo{year}{2009}).

\bibitem[{\citenamefont{Ho et~al.}(2009)\citenamefont{Ho, Cazalilla, and
  Giamarchi}}]{Ho:2009}
\bibinfo{author}{\bibfnamefont{A.~F.} \bibnamefont{Ho}},
  \bibinfo{author}{\bibfnamefont{M.~A.} \bibnamefont{Cazalilla}},
  \bibnamefont{and}
  \bibinfo{author}{\bibfnamefont{T.}~\bibnamefont{Giamarchi}},
  \bibinfo{journal}{Phys. Rev. A} \textbf{\bibinfo{volume}{79}},
  \bibinfo{eid}{033620} (pages~\bibinfo{numpages}{11}) (\bibinfo{year}{2009}).

\bibitem[{\citenamefont{F{\"o}lling et~al.}(2007)\citenamefont{F{\"o}lling,
  Trotzky, Cheinet, Feld, Saers, Widera, M\"uller, and Bloch}}]{Foelling:2007}
\bibinfo{author}{\bibfnamefont{S.}~\bibnamefont{F{\"o}lling}},
  \bibinfo{author}{\bibfnamefont{S.}~\bibnamefont{Trotzky}},
  \bibinfo{author}{\bibfnamefont{P.}~\bibnamefont{Cheinet}},
  \bibinfo{author}{\bibfnamefont{M.}~\bibnamefont{Feld}},
  \bibinfo{author}{\bibfnamefont{R.}~\bibnamefont{Saers}},
  \bibinfo{author}{\bibfnamefont{A.}~\bibnamefont{Widera}},
  \bibinfo{author}{\bibfnamefont{T.}~\bibnamefont{M\"uller}}, \bibnamefont{and}
  \bibinfo{author}{\bibfnamefont{I.}~\bibnamefont{Bloch}},
  \bibinfo{journal}{Nature} \textbf{\bibinfo{volume}{448}},
  \bibinfo{pages}{1029} (\bibinfo{year}{2007}).

\bibitem[{\citenamefont{Regal et~al.}(2003)\citenamefont{Regal, Ticknor, Bohn,
  and Jin}}]{Regal:2003b}
\bibinfo{author}{\bibfnamefont{C.~A.} \bibnamefont{Regal}},
  \bibinfo{author}{\bibfnamefont{C.}~\bibnamefont{Ticknor}},
  \bibinfo{author}{\bibfnamefont{J.~L.} \bibnamefont{Bohn}}, \bibnamefont{and}
  \bibinfo{author}{\bibfnamefont{D.~S.} \bibnamefont{Jin}},
  \bibinfo{journal}{Nature} \textbf{\bibinfo{volume}{424}}, \bibinfo{pages}{47}
  (\bibinfo{year}{2003}).

\bibitem[{\citenamefont{Andrews et~al.}(1996)\citenamefont{Andrews, Mewes,
  Druten, Kurn, and Ketterle}}]{Andrews:1996a}
\bibinfo{author}{\bibfnamefont{M.}~\bibnamefont{Andrews}},
  \bibinfo{author}{\bibfnamefont{M.-O.} \bibnamefont{Mewes}},
  \bibinfo{author}{\bibfnamefont{N.~V.} \bibnamefont{Druten}},
  \bibinfo{author}{\bibfnamefont{D.}~\bibnamefont{Kurn}}, \bibnamefont{and}
  \bibinfo{author}{\bibfnamefont{W.}~\bibnamefont{Ketterle}},
  \bibinfo{journal}{Science} \textbf{\bibinfo{volume}{273}},
  \bibinfo{pages}{84} (\bibinfo{year}{1996}).

\bibitem[{\citenamefont{St{\"o}ferle et~al.}(2006)\citenamefont{St{\"o}ferle,
  Moritz, G{\"u}nter, K{\"o}hl, and Esslinger}}]{Stoferle:2006}
\bibinfo{author}{\bibfnamefont{T.}~\bibnamefont{St{\"o}ferle}},
  \bibinfo{author}{\bibfnamefont{H.}~\bibnamefont{Moritz}},
  \bibinfo{author}{\bibfnamefont{K.}~\bibnamefont{G{\"u}nter}},
  \bibinfo{author}{\bibfnamefont{M.}~\bibnamefont{K{\"o}hl}}, \bibnamefont{and}
  \bibinfo{author}{\bibfnamefont{T.}~\bibnamefont{Esslinger}},
  \bibinfo{journal}{Phys. Rev. Lett.} \textbf{\bibinfo{volume}{96}},
  \bibinfo{pages}{030401} (\bibinfo{year}{2006}).

\bibitem[{\citenamefont{Metzner}(1991)}]{Metzner:1991}
\bibinfo{author}{\bibfnamefont{W.}~\bibnamefont{Metzner}},
  \bibinfo{journal}{Phys. Rev. B} \textbf{\bibinfo{volume}{43}},
  \bibinfo{pages}{8549} (\bibinfo{year}{1991}).

\bibitem[{\citenamefont{Henderson et~al.}(1992)\citenamefont{Henderson, Oitmaa,
  and Ashley}}]{Henderson:1992}
\bibinfo{author}{\bibfnamefont{J.~A.} \bibnamefont{Henderson}},
  \bibinfo{author}{\bibfnamefont{J.}~\bibnamefont{Oitmaa}}, \bibnamefont{and}
  \bibinfo{author}{\bibfnamefont{M.~C.~B.} \bibnamefont{Ashley}},
  \bibinfo{journal}{Phys. Rev. B} \textbf{\bibinfo{volume}{46}},
  \bibinfo{pages}{6328} (\bibinfo{year}{1992}).

\bibitem[{\citenamefont{Scarola et~al.}(2009)\citenamefont{Scarola, Pollet,
  Oitmaa, and Troyer}}]{Scarola:2009}
\bibinfo{author}{\bibfnamefont{V.~W.} \bibnamefont{Scarola}},
  \bibinfo{author}{\bibfnamefont{L.}~\bibnamefont{Pollet}},
  \bibinfo{author}{\bibfnamefont{J.}~\bibnamefont{Oitmaa}}, \bibnamefont{and}
  \bibinfo{author}{\bibfnamefont{M.}~\bibnamefont{Troyer}},
  \bibinfo{journal}{Phys. Rev. Lett.} \textbf{\bibinfo{volume}{102}},
  \bibinfo{eid}{135302} (\bibinfo{year}{2009}).

\bibitem[{\citenamefont{Partridge et~al.}(2006)\citenamefont{Partridge, Li,
  Kumar, Liao, and Hulet}}]{Partridge:2006}
\bibinfo{author}{\bibfnamefont{G.~B.} \bibnamefont{Partridge}},
  \bibinfo{author}{\bibfnamefont{W.}~\bibnamefont{Li}},
  \bibinfo{author}{\bibfnamefont{L.}~\bibnamefont{Kumar}},
  \bibinfo{author}{\bibfnamefont{Y.}~\bibnamefont{Liao}}, \bibnamefont{and}
  \bibinfo{author}{\bibfnamefont{R.~G.} \bibnamefont{Hulet}},
  \bibinfo{journal}{Science} \textbf{\bibinfo{volume}{311}},
  \bibinfo{pages}{503} (\bibinfo{year}{2006}).

\bibitem[{\citenamefont{Zwierlein et~al.}(2006)\citenamefont{Zwierlein,
  Schirotzek, Schunk, and Ketterle}}]{Zwierlein:2006}
\bibinfo{author}{\bibfnamefont{M.~W.} \bibnamefont{Zwierlein}},
  \bibinfo{author}{\bibfnamefont{A.}~\bibnamefont{Schirotzek}},
  \bibinfo{author}{\bibfnamefont{C.~H.} \bibnamefont{Schunk}},
  \bibnamefont{and} \bibinfo{author}{\bibfnamefont{W.}~\bibnamefont{Ketterle}},
  \bibinfo{journal}{Science} \textbf{\bibinfo{volume}{311}},
  \bibinfo{pages}{492} (\bibinfo{year}{2006}).

\bibitem[{\citenamefont{Rost et~al.}(2009)\citenamefont{Rost, Perry, Mercure,
  Mackenzie, and Grigera}}]{Rost:2009}
\bibinfo{author}{\bibfnamefont{A.~W.} \bibnamefont{Rost}},
  \bibinfo{author}{\bibfnamefont{R.~S.} \bibnamefont{Perry}},
  \bibinfo{author}{\bibfnamefont{J.-F.} \bibnamefont{Mercure}},
  \bibinfo{author}{\bibfnamefont{A.~P.} \bibnamefont{Mackenzie}},
  \bibnamefont{and} \bibinfo{author}{\bibfnamefont{S.~A.}
  \bibnamefont{Grigera}}, \bibinfo{journal}{Science}
  \textbf{\bibinfo{volume}{325}}, \bibinfo{pages}{1360} (\bibinfo{year}{2009}).

\bibitem[{\citenamefont{Nayak et~al.}(2008)\citenamefont{Nayak, Simon, Stern,
  Freedman, and Sarma}}]{Nayak:2008}
\bibinfo{author}{\bibfnamefont{C.}~\bibnamefont{Nayak}},
  \bibinfo{author}{\bibfnamefont{S.~H.} \bibnamefont{Simon}},
  \bibinfo{author}{\bibfnamefont{A.}~\bibnamefont{Stern}},
  \bibinfo{author}{\bibfnamefont{M.}~\bibnamefont{Freedman}}, \bibnamefont{and}
  \bibinfo{author}{\bibfnamefont{S.~D.} \bibnamefont{Sarma}},
  \bibinfo{journal}{Rev. Mod. Phys.} \textbf{\bibinfo{volume}{80}},
  \bibinfo{eid}{1083} (\bibinfo{year}{2008}).

\bibitem[{\citenamefont{Cooper and Stern}(2009)}]{Cooper:2009}
\bibinfo{author}{\bibfnamefont{N.~R.} \bibnamefont{Cooper}} \bibnamefont{and}
  \bibinfo{author}{\bibfnamefont{A.}~\bibnamefont{Stern}},
  \bibinfo{journal}{Phys. Rev. Lett.} \textbf{\bibinfo{volume}{102}},
  \bibinfo{eid}{176807} (\bibinfo{year}{2009}).

\end{thebibliography}

\end{document}